\pdfoutput=1
\documentclass[aps,prl,floatfix,superscriptaddress,twocolumn,reprint]{revtex4-1}
\usepackage{eurosym}
\usepackage{amsbsy}
\usepackage{latexsym,epsfig,graphicx}
\usepackage{dcolumn}
\usepackage{subfigure}
\usepackage{comment}
\usepackage{color}
\usepackage{bm}
\usepackage{mathrsfs}
\usepackage{amssymb}
\usepackage{amsfonts}
\usepackage{amsmath}
\usepackage{xspace}
\usepackage{epstopdf}
\usepackage{tabularx}
\usepackage{longtable}
\usepackage[colorlinks=true, letterpaper=true, pdfstartview=FitV, linkcolor=red, citecolor=blue, urlcolor=blue]{hyperref}
\usepackage[normalem]{ulem}
\usepackage[version=4]{mhchem}

\begin{document}
\title{Dynamical Singularities of Floquet Higher-Order Topological Insulators}
\author{Haiping Hu}
\affiliation{Department of Physics and Astronomy, George Mason University, Fairfax, Virginia 22030, USA}
\affiliation{Department of Physics and Astronomy, University of Pittsburgh, Pittsburgh, Pennsylvania 15260, USA}
\author{Biao Huang}
\affiliation{Department of Physics and Astronomy, University of Pittsburgh, Pittsburgh, Pennsylvania 15260, USA}
\author{Erhai Zhao}
\affiliation{Department of Physics and Astronomy, George Mason University, Fairfax, Virginia 22030, USA}
\affiliation{Quantum Materials Center, George Mason University, Fairfax, Virginia 22030, USA}
\author{W. Vincent Liu}
\affiliation{Department of Physics and Astronomy, University of Pittsburgh, Pittsburgh, Pennsylvania 15260, USA}
\affiliation{Wilczek Quantum Center, School of Physics and Astronomy and T. D. Lee Institute,  Shanghai Jiao Tong University, Shanghai 200240, China}
\affiliation{Shenzhen Institute for Quantum Science and Engineering and Department of Physics, Southern University of Science and Technology, Shenzhen 518055, China}
\begin{abstract}
We propose a versatile framework to dynamically generate Floquet higher-order topological insulators by multi-step driving of topologically trivial Hamiltonians. Two analytically solvable examples are used to illustrate this procedure to yield Floquet quadrupole and octupole insulators with zero- and/or $\pi$-corner modes protected by mirror symmetries. Furthermore, we introduce dynamical topological invariants from the full unitary return map and show its phase bands contain Weyl singularities whose topological charges form dynamical multipole moments in the Brillouin zone. Combining them with the topological index of Floquet Hamiltonian gives a pair of $\mathbb{Z}_2$ invariant $\nu_0$ and $\nu_\pi$ which fully characterize the higher-order topology and predict the appearance of zero- and $\pi$-corner modes. Our work establishes a systematic route to construct and characterize Floquet higher-order topological phases.
\end{abstract}
\maketitle
{\color{blue}\textit{Introduction.}} Topological phases of matter \cite{review1,review2} are characterized by bulk topological invariants and the appearance of robust edge/surface states. Recently, the notion of topological phases and bulk-edge correspondence has been extended to higher-order topological insulators (HOTIs) \cite{hoti01,hoti02}. A defining characteristic of HOTI is the emergence of corner or hinge modes, i.e. excitations at the intersections of edges or surfaces with energies inside the bulk gap and protected by crystalline symmetries \cite{hoti01,hoti02,hoti1,hoti2,hoti3,hoti4,hoti5,hoti6,hoti7,hoti8,hoti9,hoti10,hoti11,hoti12,hoti13,hoti14}. Theoretical concepts such as the nested Wilson loops \cite{hoti01,hoti02} and many-body multipole operators \cite{multipole1,multipole2} have been proposed to capture their topological properties and the bulk-corner/hinge correspondence. Experimentally HOTIs have been observed in phononic \cite{hotiexp1} and photonic systems \cite{hotiphotonicexp1,hotiphotonicexp2,hotiphotonicexp3}, circuit arrays \cite{hotiexp3} and crystal solids \cite{hotiexp4}.

The notion of topological phases has also been generalized to Floquet systems where the Hamiltonian is periodic in time, $H(t+T)=H(t)$, with $T$ the driving period \cite{fti1,jiangliang,lindner,kitagawa,ano1}. Periodic driving provides a powerful tool to engineer the quasienergy band structure by tuning the driving amplitude, frequency and shape. Despite the apparent similarity between quasienergy and energy, the topological properties of Floquet systems are much richer than static systems. One of its unique features is the appearance of in-gap modes pinned at quasienergy $\varepsilon=0, {\pi}/{T}$ and localized at the edge, even though the bulk quasienergy bands are trivial. Such anomalous Floquet topological insulators are intrinsically dynamical phases. In order to systematically classify Floquet topological phases \cite{fclass1,fclass2,fclass3}, one must examine the full time-evolution operator $U(t)$. In particular, the so-called return map $\tilde{U}(t)$ [see Eq. \eqref{return} below] defines a $\mathbb{Z}$ or $\mathbb{Z}_2$ topological invariant \cite{fclass2,fclass3} for each quasienergy gap. In 2D, for example, it corresponds to the winding number \cite{ano1,counter,harmonic} which counts the topological charge of Weyl-like singularities \cite{ano2,anatomy} in the instantaneous phase band during time evolution. The return map, together with the effective Hamiltonian $H_F$, can describe a large class of first-order Floquet topological insulators \cite{fclass1,fclass2,fclass3}.

It is then natural to ask whether periodic driving can give rise to new high-order topological phenomena that have no static analogues, and if so, how to characterize them? Recently, several specific models have appeared to realize Floquet HOTIs (FHOTIs) in periodically driven systems \cite{fhoti1,fhoti2,fhoti3,fhoti4,fhoti5}. These proposals however rely on building-block Hamiltonians with specific lattice structures or symmetries and are therefore not general. Moreover, the existing topological invariants in Refs. \cite{fhoti1,fhoti2,fhoti3,fhoti4,fhoti5} are supplied in a case by case manner, applicable only to a certain specific model or symmetry class. A theory for FHOTIs that can predict the corner modes (CMs) from bulk invariants constructed from a general $\tilde{U}(t)$ and $H_F$ is still lacking.

Motivated by these considerations, in this paper we demonstrate a generic route to realize and characterize FHOTIs. The construction does not rely on any specific space-time symmetries of the building-block Hamiltonians. As an example, a 2D model is solved analytically to determine the phase diagram, which contains two Floquet quadrupole topological phases with 0- and $\pi$-CMs respectively. Via the decomposition of the unitary evolution, we show that the topology of the quasienergy bands is captured by $\mathbb{Z}_2$ invariant $\nu^F_0$ from the nested Wilson loops, while the return maps feature {\it multipole patterns of dynamical singularities}: the topological charges of the Weyl-type singularities of $\tilde{U}(t)$ form a quadrupole moment in the Brillouin zone (BZ) at certain instants. Two dynamical invariants $n_0, n_{\pi}$ are introduced to count these charges. From $\nu_0^F$ and $n_{0,\pi}$, we show that each quasienergy gap is characterized by a $\mathbb{Z}_2$ index $\nu$ that predicts the appearance or absence of CMs. The new $\mathbb{Z}_2$ invariants work for all mirror-symmetry protected FHOTIs and go beyond the periodic table of first-order Floquet topological insulators. The construction and topological analysis are then generalized to 3D Floquet octupole topological insulators.

{\color{blue}\textit{Dynamical construction of FHOTI.}} The dynamics of a periodically driven lattice system with Hamiltonian $H(t)$ is governed by the unitary evolution $U(t)=\mathcal{T}e^{-i\int _0^t H(\tau)d\tau}$, where $\hbar=1$ and $\mathcal{T}$ denotes time-ordering. To extract its topology, it is convenient to decompose $U(t)$ into a unitary loop $\tilde{U}(t)$ satisfying $\tilde{U}(0)=\tilde{U}(T)=I$ and the time evolution of a constant Hamiltonian $H_F$ \cite{fclass2}. Explicitly, one can define the effective Hamiltonian $H_F=i\log U(T)/T$ as well as the return map \cite{ano1,fclass2,fclass3}
\begin{eqnarray}
\tilde{U}(t) =U(t)e^{i H_Ft}.\label{return}
\end{eqnarray}
Usually, $\tilde{U}(t)$ is defined for a given gap with the logarithm branch cut lying within it. It is apparent from Eq. (\ref{return}) that the topology of $U(t)$ is carried by both $H_F$ and $\tilde{U}(t)$. The spectra $\varepsilon_n$ of $H_F$ are known as quasienergy bands and we take $\varepsilon_n\in [-\pi/T, \pi/T]$.
\begin{figure}[!t]
\centering
\includegraphics[width=3.35in]{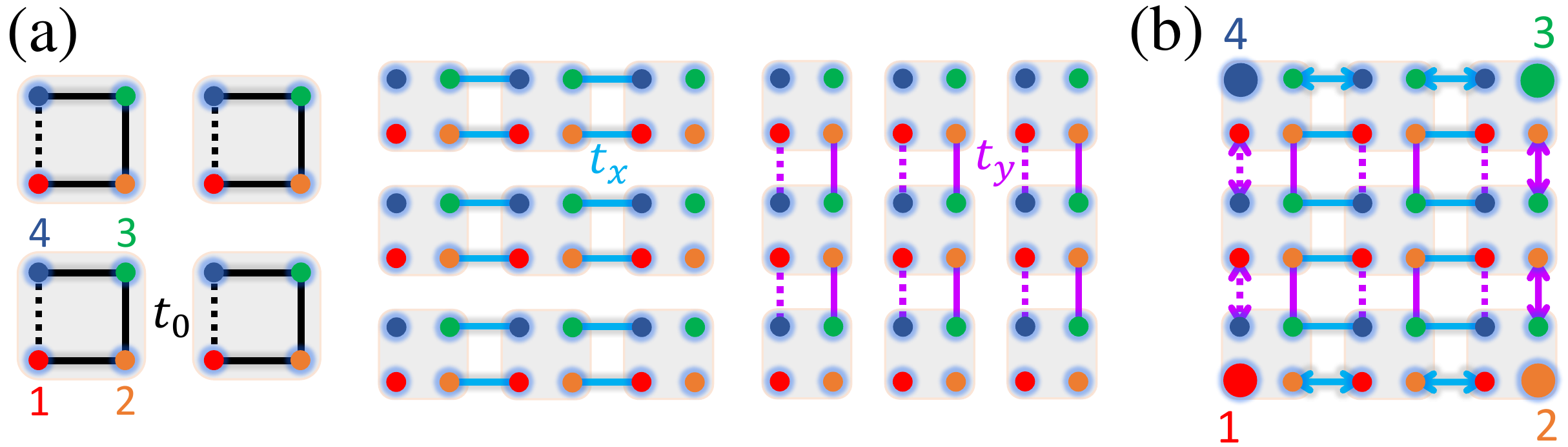}
\caption{Construction of FHOTI on a square lattice from multi-step driving. (a) The trivial building blocks $h_0$ (left), $h_x$ (middle), and $h_y$ (right) with intra-cell hopping $t_0$ and inter-cell hopping $t_x, t_y$. Dashed lines represent hoppings with negative signs. (b) Schematic of particle motion in one period of two-step driving, $h_y$ followed by $h_x$ with $t_0=0$. The corners are dynamically decoupled from the bulk, giving rise to four localized corner modes (big solid circles).}
\label{fig1}
\end{figure}

The basic idea of dynamical construction of FHOTI can be illustrated by a simple example of Floquet quadrupole insulator depicted in Fig. \ref{fig1}(a). Consider a square lattice, where each unit cell (shaded box) consists of four lattice sites. Our strategy is to herd the motion (more precisely the quantum walks) of particles by spatial control of the tunneling amplitudes in multiple steps  within each driving period. Three trivial Hamiltonians $h_x$, $h_y$ and $h_0$ serve as the building blocks: $h_{x/y}$ only contains inter-cell hopping $t_{x/y}$ along the $x/y$ direction, and $h_0$ only contains intra-cell hopping $t_0$. To visualize the emergence of topological CMs, consider the limit of $t_0=0$ and two-step driving: $H(t<T/2)=h_y$ followed by $H(t>T/2)=h_x$. The semiclassical particle motion is sketched  in Fig. \ref{fig1}(b). It is clear that particles in the bulk move along a plaquette, while particles on the four edges hop back and forth. However, particles initially at the four corners remain localized and completely decoupled from the bulk and edge dynamics. They are nothing but Floquet CMs. We will show below that the CMs persist to finite $t_0$ as the bulk excitations form Floquet bands separated by gaps. Similar to static case \cite{hoti01,hoti02}, the Floquet CM is protected by crystalline symmetries (e.g., mirror reflection).

This picture motivates us to propose the following generic $N$-step driving sequence. In each step $s$ with time interval $T_s$, the system evolves according to a constant Hamiltonian $h_s$ assumed, for simplicity, to be a sum of anti-commuting terms (see $h_0$, $h_{x,y}$ in Eq. \eqref{driving} below). Accordingly,
\begin{eqnarray}
U(T)=\prod\nolimits_{s=1}^{N} (\cos\theta_s-i\sin\theta_s \tilde{h}_s).\label{nstep}
\end{eqnarray}
Here $\theta_s=T_s|E_s|$, $\tilde{h}_s=h_s/|E_s|$, with $\pm E_s$ the spectrum of $h_s$. By definition, the wave functions of CMs at quasienergy zero (0-CM) and $\pi/T$ ($\pi$-CM) satisfy
\begin{eqnarray}
U(T)|\psi_0\rangle=|\psi_0\rangle,~~~U(T)|\psi_{\pi}\rangle=-|\psi_{\pi}\rangle.
\end{eqnarray}
The existence of solution to these eigen equations is guaranteed by properly choosing $\theta_s$ and $h_s$ as follows. Consider a state $|\eta\rangle$ localized at the corner (Fig. 1b). It may couple to neighboring sites by $h_{s=1}$ in the first step. But for all other steps $s> 1$, $h_s$ is chosen so $h_{s>1}|\eta\rangle=0$. A 0-CM is realized if we choose $\theta_1=0$. Its wave function $|\psi_{0}\rangle$ is simply given by $|\eta\rangle$. Similarly setting $\theta_1=\pi$ gives rise to $\pi$-CM with $|\psi_{\pi}\rangle=|\eta\rangle$. For 0- and $\pi-$CMs to coexist \cite{fhoti1}, one can choose for example $\theta_1=\pi/2$ and $\theta_{s>1}=\pi$ for even $N$. We will give a few examples below to illustrate how this construction procedure can be applied to generate different kinds of FHOTIs.

{\color{blue}\textit{Floquet quadrupole insulator.}} First we present an analytically solvable model of Floquet quadrupole insulator (FQI) and demonstrate the emergence of topological CMs. The overall set up has been introduced above in Fig. \ref{fig1} on the square lattice. The $2\times 2$ unit cell is conveniently described by two sets of Pauli matrices $\bm{\sigma}$ and $\bm{\tau}$. The trivial building blocks are hopping Hamiltonians $h_0=t_0(\tau_0\sigma_1+\tau_2\sigma_2)$, $h_x=t_x(\cos k_x\tau_0\sigma_1-\sin k_x\tau_3\sigma_2)$, and $h_y=t_y(\cos k_y\tau_2\sigma_2+\sin k_y\tau_1\sigma_2)$ where $\bm k=(k_x,k_y)$ is the quasi-momentum. The terms in $h_{0,x,y}$ anti-commute and the system possesses two mirror symmetries $\mathcal{M}_x=i\tau_3\sigma_1$ and $\mathcal{M}_y=i\tau_1\sigma_1$. The driving protocol is
\begin{eqnarray}
&&t\in T_1,H(t)=h_0;~~t\in T_2, H(t)=h_y;\nonumber\\
&&t\in T_3,H(t)=h_x;~~t\in T_4,H(t)=h_0,\label{driving}
\end{eqnarray}
with time interval $T_s=[(s-1)T/4,sT/4$). For $t_xT=t_yT=\pi$, the FQI phase with 0-CMs appears when \cite{SM}
\begin{eqnarray}
(\mathcal{N}-1/6)\pi<\phi_0<(\mathcal{N}+1/6)\pi,~~\mathcal{N}\in\mathbb{Z}.\label{zeropc}
\end{eqnarray}
with $\phi_0\equiv\frac{t_0T}{2\sqrt{2}}$. The FQI phase with $\pi$-CMs lies within
\begin{eqnarray}
(\mathcal{N}+1/3)\pi<\phi_0<(\mathcal{N}+2/3)\pi,~~\mathcal{N}\in\mathbb{Z}.\label{pipc}
\end{eqnarray}
For all other values of $\phi_0$, the system is a trivial band insulator with no CMs.

\begin{figure}[!t]
\centering
\includegraphics[width=3.25in]{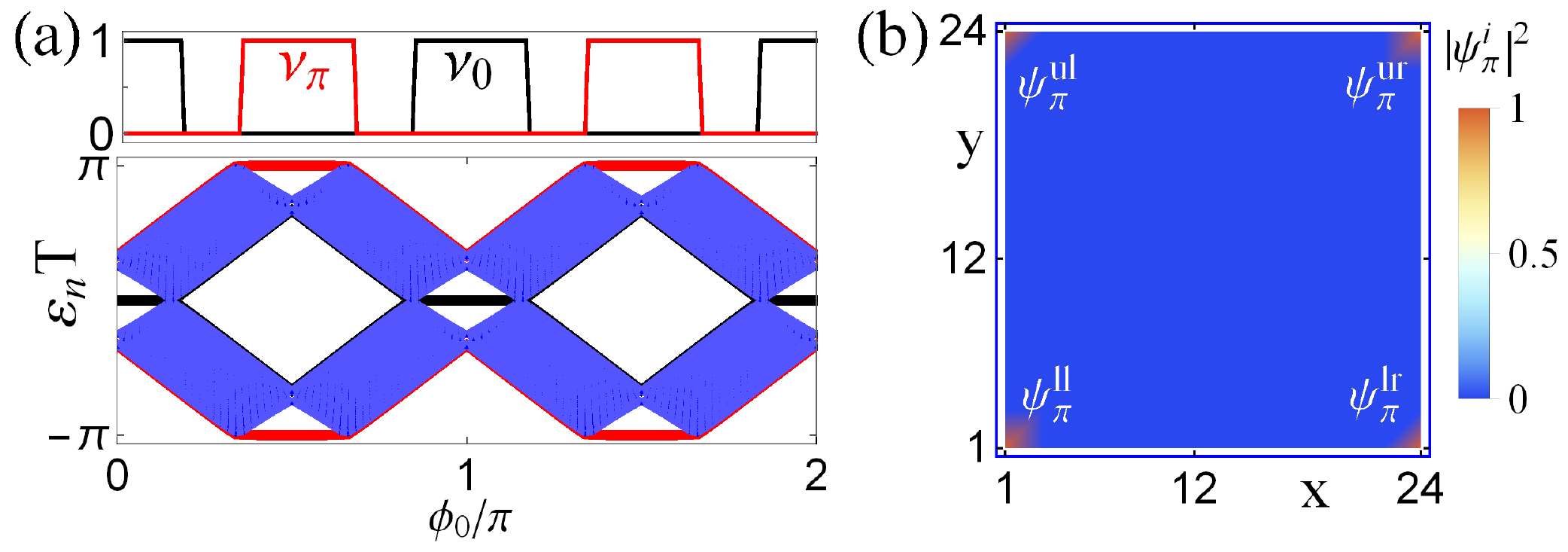}
\caption{(a) Phase diagram of the Floquet system with driving Eq. \eqref{driving}. Top: topological invariants $\nu_{0}$ (black) and $\nu_{\pi}$ (red) obtained from Eq. (\ref{topoinva}) showing two FQI phases. Bottom:
the quasienergy spectra for a finite $24\times 24$ lattice. The four-fold degenerate 0-/$\pi$-CMs are marked by the black/red lines, respectively. (b) The spatial wave functions of four $\pi$-CMs, $|\psi_{\pi}^{i}|^2(i=ll,lr,ul,ur)$, $\phi_0/\pi=0.45$, $t_x=t_y=\pi/T$.}
\label{fig2}
\end{figure}

Fig. \ref{fig2}(a) shows the quasienergy spectra as function of $\phi_0$ for a finite lattice with open boundary conditions. In between the bulk bands, we observe four-fold degenerate in-gap modes pinned at $\varepsilon=0$ or $\varepsilon=\pi/T$. They appear alternatively with a period of exactly $\pi$ as $\phi_0$ is varied, and are separated from each other by the topologically trivial phase, in consistent with Eqs. \eqref{zeropc}-\eqref{pipc}. The wave functions of these in-gap modes are shown in Fig. \ref{fig2}(b). They are indeed localized at the four corners arising from the bulk quadrupoles. In comparison, the quasienergy spectra for periodic boundary condition or stripe geometry are fully gapped \cite{SM}, indicating vanishing conventional dipoles.

This model provides an elegant example of our dynamical construction of FHOTIs and CMs summarized in Eq. \eqref{nstep}. Denote the wave functions of four CMs as $|\psi_{0/\pi}^{i}\rangle$ ($i=ll,lr,ul,ur)$ and take $i=ll$, the lower-left corner for example. For $\phi_0=0$, the 0-CM wave function is localized at a single site labeled as 1 [Fig. \ref{fig1}], $|\psi^{ll}_0\rangle=|1\rangle_{ll}$, corresponding to the value $\theta_1=0$ in our construction scheme. The other two driving steps $h_{x,y}$ do not couple the CMs to the bulk. For $\phi_0=\pi/2$, the $\pi$-CM wave function is $|\psi^{ll}_{\pi}\rangle=\frac{1}{\sqrt{2}}(|2\rangle_{ll}-|4\rangle_{ll})$, corresponding to $\theta_1=\pi$. When deviating from these ideal limits, the CMs spread further into the bulk but remain localized. The FQI and CMs persist as long as the bulk gaps stay open.

{\color{blue}\textit{Dynamical topological invariants.}} For static HOTI, the higher-order bulk topology and appearance of CMs can be described by introducing Wannier bands and nested Wilson loops \cite{hoti01,hoti02,SM,wannier}. The analysis can be generalized to Floquet systems to capture the topological properties of $H_F$ and the quasienergy bands. We chose the lower two overlapping quasienergy bands to construct the Wannier-band subspace $|\mathcal{\omega}_{x,\bm k}^{j}\rangle$ ($j=1,2$) and compute the nested polarizations \cite{hoti01,hoti02,SM}, for example,
\begin{eqnarray}
p_y^j= i\int_{BZ}\frac{d^2 k}{(2\pi)^2} \langle\mathcal{\omega}_{x,\bm k}^{j}|\partial_{k_y}|\mathcal{\omega}_{x,\bm k}^{j}\rangle.
\end{eqnarray}
In the presence of mirror symmetries $\mathcal{M}_x$ and $\mathcal{M}_y$, the nested polarization $p_y^{j}$ and $p_x^{j}$ are quantized to be $0$ (trivial) or ${1}/{2}$ (topological) \cite{hoti01,hoti02}, yielding a $\mathbb{Z}_2$ classification. The topological quadrupole phase corresponds to $(p_y^{j},p_x^j)=({1}/{2},{1}/{2})$. It is characterized by $\mathbb{Z}_2$ invariant
\begin{eqnarray}
\nu_0^F=4p_y^{{j}}p_x^{{j}}.\label{tqi}
\end{eqnarray}
For the two FQI phases above, $\nu_0^F$ is found to be $1$, which is consistent with the quantized tangential polarization along the edges \cite{hoti01,hoti02,SM}. By itself, however, $\nu_0^F$ cannot distinguish the two FQI phases, or predict in which gap the CMs reside or even the existence of CMs (e.g. for anomalous FQI \cite{fhoti1}, $\nu_0^F$ is zero but CMs are present). This is not surprising because it only captures the topology of $H_F$, not the full $U(t)$. For FHOTI, an intrinsically dynamical topological invariant is needed.
\begin{figure}[!t]
\centering
\includegraphics[width=3.25in]{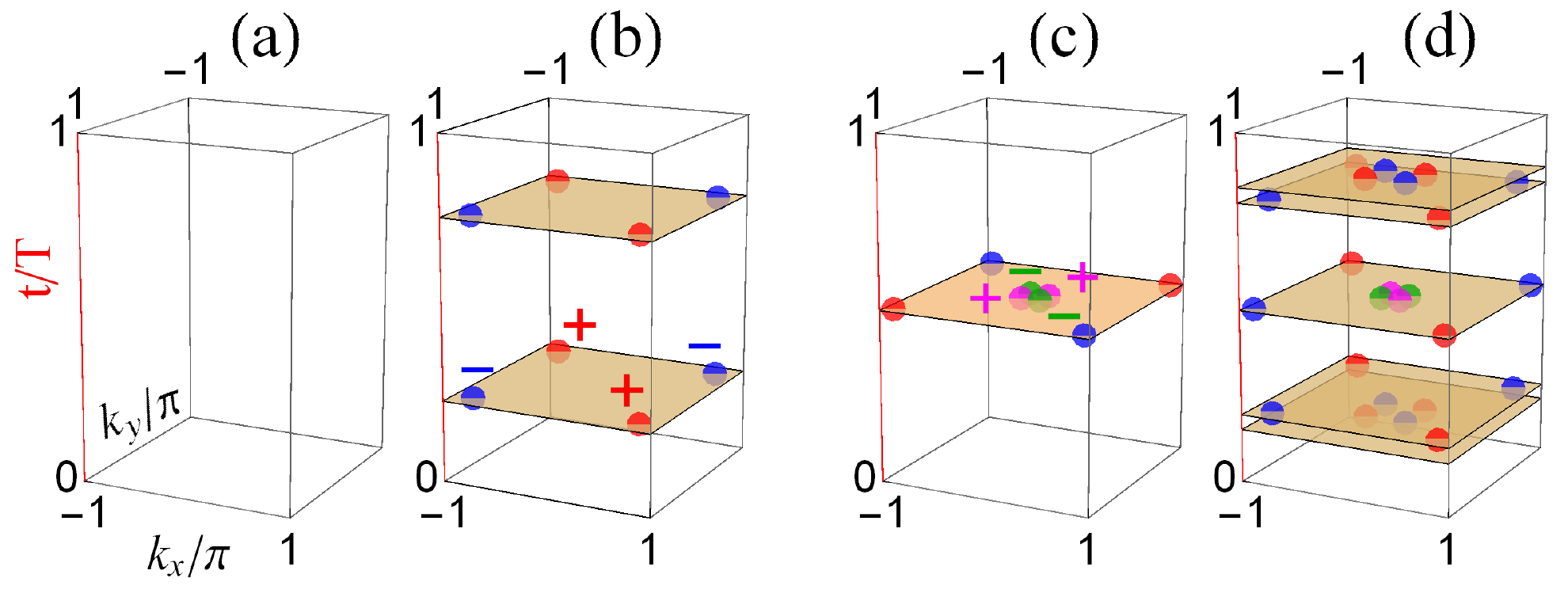}
\caption{Dynamical singularities of FQIs in $(\bm k,t)$-space. The colored dots label the Weyl charges in the phase band of $\tilde{U}(t)$ at certain time instants. Their topological charges form quadrupole moments in the BZ. The red/blue dots label charge $\pm 1$ at the $\pi$-gap; the magenta/green dots label charge $\pm 1$ at the $0$-gap. From (a) to (d), $\phi_0/\pi=0.1$, $0.95$, $0.45$, $1.45$ with $t_x=t_y=\pi/T$.}
\label{fig3}
\end{figure}

Such a dynamical invariant can be defined from the return map $\tilde{U}(t)$. The diagonalization of $\tilde{U}$ yields $\tilde{U}(t)=\sum_m e^{-i\tilde{\varepsilon}_m(\bm k,t)}|\varphi_m(\bm k,t)\rangle\langle\varphi_m(\bm k,t)|$, with the eigenphases $\tilde{\varepsilon}_m$ forming the phase bands \cite{ano2,fclass2}. For our system, during the time evolution $t\in(0,T)$, the gap may close at $0$ or $\pi/T$ as the phase bands touch each other at isolated points in the $(\bm k,t)$-space, similar to Weyl points in semimetals, and reopen afterwards. Such singular points resemble magnetic monopoles and carry topological charges \cite{ano2}. For the $i$-th degeneracy point $\bm d_j=(\bm k_j,t_j)$ of band $m$, we compute its topological charge $C_j=\frac{1}{2\pi i}\oint_{\bm S_j}\bm{\nabla}\times\langle\varphi_m(\bm k,t)|\bm{\nabla}|\varphi_m(\bm k,t)\rangle\cdot d\bm S$, with $\bm{S}_j$ a small surface enclosing $\bm d_j$.

Due to the mirror symmetries $\mathcal{M}_{x,y}$, these ``Weyl points'' at a specific time instant always come in quartets, i.e. at $\bm k=(\pm k_x,\pm k_y)$ in the 2D BZ. And their charges form a quadrupole pattern \cite{footnote} as illustrated in Fig. \ref{fig3}(a)-(d). Such dynamical quadrupole (with zero total charge) indicates the higher-order topology and the absence of 1D edge states \cite{ano1,ano2}.  In fact, one can prove that a quadrupole pattern is equivalent to its mirror image by a
continuous deformation based on $\mathcal{M}_x$ or $\mathcal{M}_y$ \cite{SM}. Thus, $n_{0,\pi}=\sum_{\bm k_j\in\textrm{1st qBZ}}^{\tilde{\varepsilon}(\bm d_j)=0,\pi}C_j$, the total Weyl charge within the first quadrant of the BZ during $t\in(0,T)$, is only defined modulo 2, and its parity can serve as the dynamical invariant for corresponding gap.
Combining $n_{0,\pi}$ from $\tilde{U}(t)$ with the quadrupole invariant $\nu_0^F$ for $H_F$, we arrive at two $\mathbb{Z}_2$-valued
invariants $\nu_{0,\pi}$ for the 0- and $\pi$-gap respectively (for details, see \cite{SM}),
\begin{eqnarray}
\nu_\pi=n_{\pi} \bmod 2;~~~\nu_0=(n_0+\nu_0^F) \bmod 2.\label{topoinva}
\end{eqnarray}
We stress that the $\mathbb{Z}_2$ nature of $\nu_{0,\pi}$ originates from mirror symmetries. A nonzero value of $\nu_{0}=1$ ($\nu_{\pi}=1$) indicates the appearance of CMs at the $0$-gap ($\pi$-gap). Thus, our Floquet system follows a $\mathbb{Z}_2\times\mathbb{Z}_2$ classification and is described by two $\mathbb{Z}_2$ invariants ($\nu_0,\nu_{\pi}$), one for each gap. To check the correspondence between bulk invariants Eq. \eqref{topoinva} and the CMs observed in numerics, we give a few examples of the Weyl charges in Figs. \ref{fig3}(a)-(d). For the FQI phase with 0-CMs [Fig. \ref{fig3}(a)(b)], we have $n_0=0$ and $n_{\pi}=0$ or $2$. In both cases, $(\nu_0,\nu_{\pi})=(1,0)$. For the FQI phase with $\pi$-CMs [Fig. \ref{fig3}(c)(d)], $n_0=1$ and $n_{\pi}=1$ or $5$. Thus, $(\nu_0,\nu_{\pi})=(0,1)$. It is clear that Eq. (\ref{topoinva}) correctly predicts the appearance of Floquet CMs, in agreement with Fig. \ref{fig2}(a). We have checked that the invariants $\nu_{0,\pi}$ also apply to anomalous FQIs with $\nu_0=\nu_{\pi}=1$ discussed in \cite{fhoti1,SM}.

\begin{figure}[!t]
\centering
\includegraphics[width=3.4in]{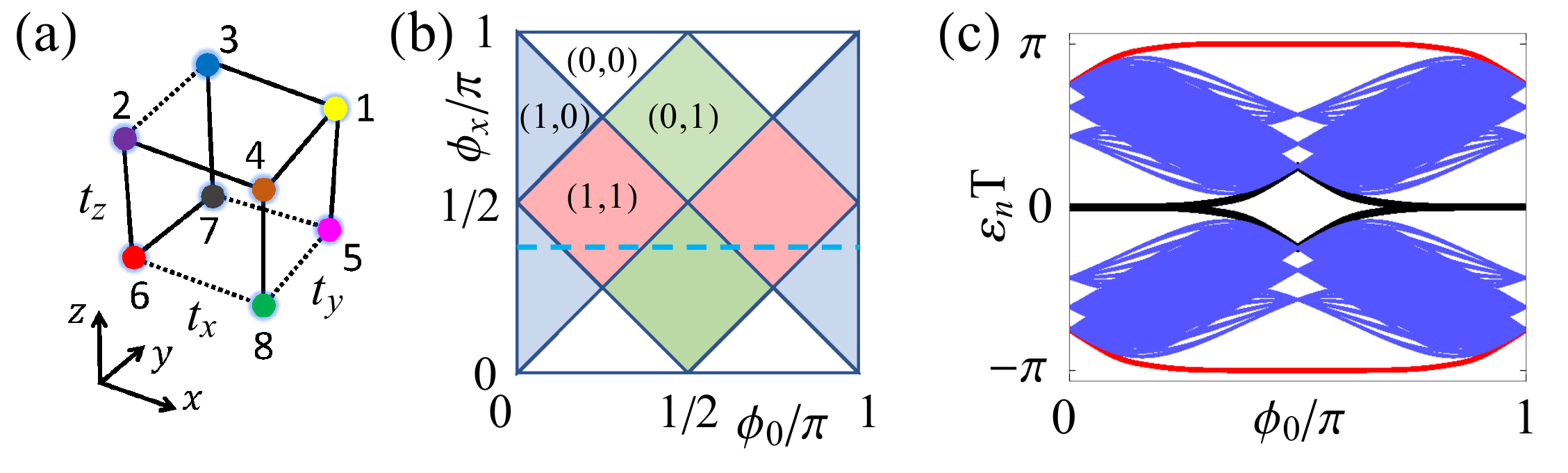}
\caption{Floquet octupole insulator. (a) The unit cell contains 8 sites on a cubic lattice, the solid/dash lines denote hoppings with $+/-$ signs. (b) The phase diagram with $\phi_0$ and $\phi_x$ defined in the main text. Color-coded regions represent three FOI phases with 0-CMs only (blue), $\pi$-CMs only (green), both 0- and $\pi$-CMs (red), and the trivial phase (white). Each phase is labeled by its dynamical invariants $(\nu_0,\nu_{\pi})$. (c) Quasienergy spectra of a $16\times16\times16$ lattice along the dash line in (b) for fixed $\phi_x/\pi=3/8$. The black/red lines mark the eight-fold degenerate $0/\pi$-CMs.}
\label{octupole}
\end{figure}

{\color{blue}\textit{Floquet octupole insulator.}} Next we show how to generate Floquet octupole insulators (FOIs) on a cubic lattice following our general scheme. The degrees of freedom inside the eight-site unit cell, illustrated in Fig. \ref{octupole}(a), can be described by three sets of Pauli matrices $\bm{\tau}$, $\bm{\sigma}$ and $\bm{s}$. The dynamical construction employs four building blocks: an intra-unit cell hopping Hamiltonian $h_0=t_0(\Gamma_2+\Gamma_4+\Gamma_6)$ and three inter-unit cell hopping Hamiltonians $h_x=t_x(\sin k_x\Gamma_3+\cos k_x\Gamma_6)$, $h_y=t_y(\sin k_y\Gamma_1+\cos k_y\Gamma_2)$, $h_z=t_z(\sin k_z\Gamma_5+\cos k_z\Gamma_4)$ with $\Gamma_0=\tau_3\sigma_3 s_0$, $\Gamma_i=-\tau_3\sigma_2 s_i$ for $i=1,2,3$, $\Gamma_4=\tau_1\sigma_0 s_0$, $\Gamma_5=\tau_2\sigma_0 s_0$ and $\Gamma_6=i\prod_{j=0}^5\Gamma_j$. The driving protocol consist of two steps: for $0<t<T/4$ and $3T/4<t<T$, $H(t)=h_0$; for $T/4<t<3T/4$, $H(t)=h_x+h_y+h_z$. Let us focus on the simple case of $t_x=t_y=t_z$. Then the phase boundaries can be found analytically \cite{SM},
\begin{eqnarray}
\phi_0\pm\phi_x=\mathcal{N}\pi/2,~~\mathcal{N}\in \mathbb{Z}\label{pb2}
\end{eqnarray}
with $\phi_0=\sqrt{3}t_0T/4$ and $\phi_x=\sqrt{3}t_xT/4$.

The phase diagram on the $\phi_0-\phi_x$ plane is depicted in Fig. \ref{octupole}(b). It contains three distinct FOIs and a trivial phase. Roughly speaking, the FOI phase with only 0-CMs is located near $\phi_0=0$ and $\pi$, while the FOI phase with only $\pi$-CMs occupies regions around $\phi_0=\pi/2$. Sandwiched in between is the third, anomalous FOI which has both 0- and $\pi$-CMs. The quasienergy spectrum for a finite system with open boundary conditions is shown in Fig. \ref{octupole}(c) for parameters along a cut in the phase diagram with fixed $\phi_x=3\pi/8$. The location of different Floquet CMs agrees with the phase boundaries given by Eq. (\ref{pb2}). To cast this example in the general scheme Eq. \eqref{nstep}, we notice the 0-CM at point $\phi_0=0$ is simply $|\psi_{0}\rangle=|6\rangle$ with $\theta_1=0$. The $\pi$-CM at $\phi_0=\pi/2$ is just $|\psi_{\pi}\rangle=(|2\rangle+|7\rangle-|8\rangle)/{\sqrt{3}}$ with $\theta_1=\pi$. The system has three mirror symmetries: $\mathcal{M}_x=\tau_0\sigma_1 s_3$; $\mathcal{M}_y=\tau_0\sigma_1 s_1$; and $\mathcal{M}_z=\tau_0\sigma_3 s_0$. Together they quantize the octupole moment. Similar to the FQIs, the topology of the Floquet system is carried by both $H_F$ and the return map $\tilde{U}(t)$. The former is characterized by a $\mathbb{Z}_2$ invariant $\nu_0^F$ \cite{SM}; the latter contains singularities of the phase bands in 4D $(\bm k,t)$-space. We find the invariants in Eq. (\ref{topoinva}) are still valid  \cite{SM}.

{\color{blue}\textit{Outlook.}} We have introduced a versatile route to construct and characterize FHOTIs. The building blocks are topologically trivial and accessible in many synthetic (e.g. photonic and cold-atoms) quantum systems. For example, the quadrupole phase can be realized based on the $\pi$-flux Hofstadter model \cite{piflux1,piflux2} with the addition of a superimposed superlattice along both the $x$ and $y$ directions \cite{hoti01}. Alternatively, the modulation along one direction may be replaced by utilizing spin degree of freedom, with the effective hoppings being induced by Raman coupling and laser-assisted tunneling in different directions, respectively. The driving protocol can be viewed more generally as discrete-time quantum walks on lattice \cite{qw1,qw2,qw3}. By imposing further constraints on the building blocks or the driving protocols, our construction can be generalized to realize higher-order topological phases in other symmetry classes. In contrast to previous constructions of model-dependent topological invariants, the phase-band singularities are general for Floquet systems, hinting the possibility of a unified scheme for characterizing the higher-order topology for a wide class of systems. Experimentally, in addition to the observation of CMs, the higher-order topology may be identified from the tomography of band-touching singularities \cite{btproposal}. Finally, it would be interesting to investigate FHOTIs in the frequency domain \cite{ano1,frequency} and the time evolution of CMs from the entanglement perspective \cite{SM}.

\begin{acknowledgments}
This work is supported by NSF Grant No. PHY-1707484 (H.H. and E.Z.), AFOSR Grant No.  FA9550-16-1-0006 (H.H., E.Z., and W.V.L.), MURI-ARO Grant No. W911NF-17-1-0323 (B.H. and W.V.L.), and NSF of China Overseas Scholar Collaborative Program Grant No. 11429402 sponsored by Peking University (W.V.L.).
\end{acknowledgments}

\clearpage
\onecolumngrid
\appendix
\section{Supplementary materials for ``Dynamical singularities of Floquet higher-order topological insulators"}
In this supplementary material, we provide details on the derivation of stroboscopic evolution operator $U(T)$ for driving protocol Eq. (4), Floquet spectra under different boundary conditions, nested Wilson loop (NWL) approach, justification of $\mathbb{Z}_2$ invariants, phase-band characterizations for the anomalous Floquet quadrupole insulators (FQIs) and Floquet octupole insulators (FOIs), time-evolution of corner modes (CMs).

\subsection{Derivation of stroboscopic evolution operator and phase boundaries}
While the driving consists of three steps, for analytical convenience we have shifted the time origin, which amounts to a gauge transformation, and partitioned the time evolution into four parts. The stroboscopic evolution operator $U(T)$ for our driving protocol Eq. (4) can be explicitly calculated as ($t_x=t_y$ is assumed, $\hbar=1$, $\phi_x\equiv\frac{t_xT}{4}$, $\phi_y\equiv\frac{t_yT}{4}$, $\phi_0\equiv\frac{t_0T}{2\sqrt{2}}$, $\tilde{h}_0\equiv\frac{h_0}{t_0}$, $\tilde{h}_{x,y}\equiv\frac{h_{x,y}}{t_{x,y}}$).
\begin{eqnarray}
U(T)&=&e^{-i \frac{h_0T}{4}}e^{-i \frac{h_xT}{4}}e^{-i \frac{h_yT}{4}}e^{-i \frac{h_0T}{4}}\nonumber\\
=&&[\cos\phi_0-i\sin\phi_0\frac{\tilde{h}_0}{\sqrt{2}}][\cos\phi_x-i\sin\phi_x \tilde{h}_x][\cos\phi_y-i\sin\phi_y \tilde{h}_y][\cos\phi_0-i\sin\phi_0\frac{\tilde{h}_0}{\sqrt{2}}]\nonumber\\
=&&[\cos\phi_0-i\sin\phi_0\frac{\tilde{h}_0}{\sqrt{2}}][\cos^2\phi_x-\frac{i}{2}\sin 2\phi_x(\tilde{h}_x+\tilde{h}_y)-\sin^2\phi_x \tilde{h}_x \tilde{h}_y][\cos\phi_0-i\sin\phi_0\frac{\tilde{h}_0}{\sqrt{2}}]\nonumber\\
=&&\cos^2\phi_x[\cos 2\phi_0-\frac{i}{\sqrt{2}}\sin 2\phi_0 \tilde{h}_0]\notag\\
&&-\frac{i}{2}\sin 2\phi_x[\cos^2\phi_0 (\tilde{h}_x+\tilde{h}_y)-\frac{i}{\sqrt{2}}\sin 2\phi_0(\cos k_x+\cos k_y)-\frac{\sin^2\phi_0}{2} (\tilde{h}_0 \tilde{h}_x \tilde{h}_0+\tilde{h}_0 \tilde{h}_y \tilde{h}_0)]\notag\\
&&-\sin^2\phi_x[\cos^2\phi_0 \tilde{h}_x \tilde{h}_y-\frac{i}{2\sqrt{2}}\sin 2\phi_0(\tilde{h}_0 \tilde{h}_x \tilde{h}_y+\tilde{h}_x \tilde{h}_y \tilde{h}_0)-\frac{\sin^2\phi_0}{2}\tilde{h}_0 \tilde{h}_x \tilde{h}_y \tilde{h}_0)].\label{utop}
\end{eqnarray}
Each term in Eq. (\ref{utop}) can be expanded as
\begin{eqnarray}
&&\tilde{h}_0 \tilde{h}_x \tilde{h}_0/2=\sin k_x \tau_3\sigma_2+\cos k_x\tau_2\sigma_2,\nonumber\\
&&\tilde{h}_0 \tilde{h}_y \tilde{h}_0/2=-\sin k_y \tau_1\sigma_2+\cos k_y\tau_0\sigma_1,\nonumber\\
&&\tilde{h}_x \tilde{h}_y=i\cos k_x\cos k_y\tau_2\sigma_3+i\sin k_x\cos k_y\tau_1\sigma_0+i\cos k_x\sin k_y\tau_1\sigma_3-i\sin k_x\sin k_y\tau_2\sigma_0,\nonumber\\
&&\{\tilde{h}_0, \tilde{h}_x \tilde{h}_y\}/2=i\sin k_x\cos k_y\tau_1\sigma_1-i\sin k_x\sin k_y\tau_2\sigma_1-i\sin k_x\sin k_y\tau_0\sigma_2+i\cos k_x\sin k_y\tau_3\sigma_1,\nonumber\\
&&\tilde{h}_0 \tilde{h}_x \tilde{h}_y \tilde{h}_0/2=-i \sin k_x\cos k_y\tau_3\sigma_3+i\cos k_x\sin k_y\tau_3\sigma_0-i\sin k_x\sin k_y\tau_2\sigma_0-i\cos k_x\cos k_y\tau_2\sigma_3.
\end{eqnarray}
As $U(T)$ is a unitary operator, it can be rewritten as the sum of a real and imaginary Hermite operator, i.e., $U(T)=f(\bm k)+i g(\bm k)$ with $f(\bm k)=f^{\dag}(\bm k)$ and $g(\bm k)=g^{\dag}(\bm k)$. After some calculations, one finds
\begin{eqnarray}
f(\bm k)&=&\cos^2\phi_x\cos 2\phi_0-\frac{1}{2\sqrt{2}}\sin 2\phi_x\sin 2\phi_0(\cos  k_x+\cos k_y)\notag\\
&&-\frac{\sin^2\phi_x\sin 2\phi_0}{\sqrt{2}}(\sin k_x\cos k_y\tau_1\sigma_1-\sin k_x\sin k_y\tau_2\sigma_1-\sin k_x\sin k_y\tau_0\sigma_2+\cos k_x\sin k_y\tau_3\sigma_1).
\end{eqnarray}

The topological phase transitions are determined by the gap closings located at $0$ and $\frac{\pi}{T}$, i.e., when the eigenvalues of $f(\bm k)$ take $\pm 1$. For the specific case discussed in the main text, $\phi_x=\phi_y=\frac{t_xT}{4}=\frac{\pi}{4}$, the Dirac points satisfy $k_x=\pm k_y$. We have
\begin{eqnarray}
Eig[f(\bm k)]=\frac{\cos 2\phi_0}{2}-\sin 2\phi_0 (\frac{\cos k_x}{\sqrt{2}}\pm\frac{\sin k_x}{2}).
\end{eqnarray}
As $|\frac{\cos k_x}{\sqrt{2}}\pm \frac{\sin x}{2}|\leq \frac{\sqrt{3}}{2}$, the 0-gap closing condition dictating the emergence of 0-CMs is then
\begin{eqnarray}
\frac{\cos 2\phi_0}{2}\pm\frac{\sqrt{3}\sin 2\phi_0}{2}=1,
\end{eqnarray}
with solutions
\begin{eqnarray}
\phi_{0,\mathcal{N}\pm}=(\mathcal{N}\pm\frac{1}{6})\pi, ~\mathcal{N}\in \mathbb{Z}.
\end{eqnarray}
Similarly, the $\pi$-gap closing condition which dictating the emergence of $\pi$-CMs is
\begin{eqnarray}
\frac{\cos 2\phi_0}{2}\pm\frac{\sqrt{3}\sin 2\phi_0}{2}=-1,
\end{eqnarray}
with solutions
\begin{eqnarray}
\phi_{\pi,\mathcal{N}\pm}=(\mathcal{N}\pm\frac{1}{3})\pi, ~\mathcal{N}\in\mathbb{Z}.
\end{eqnarray}

\subsection{Floquet spectra under different boundary conditions}
The bulk-edge-corner correspondence manifest itself through the Floquet spectra under different boundary conditions. In Fig. 2(a) of the main text, we have demonstrated the Floquet spectra of driving protocol Eq. (4) under open boundary conditions (OBC). There we have observed the emergence of two types of FQIs, supporting either 0- or $\pi$-CMs, respectively. As a comparison, here we show their corresponding Floquet spectra under periodical boundary conditions (PBC) and stripe geometry.
\begin{figure}[!h]
\centering
\includegraphics[width=5.5in]{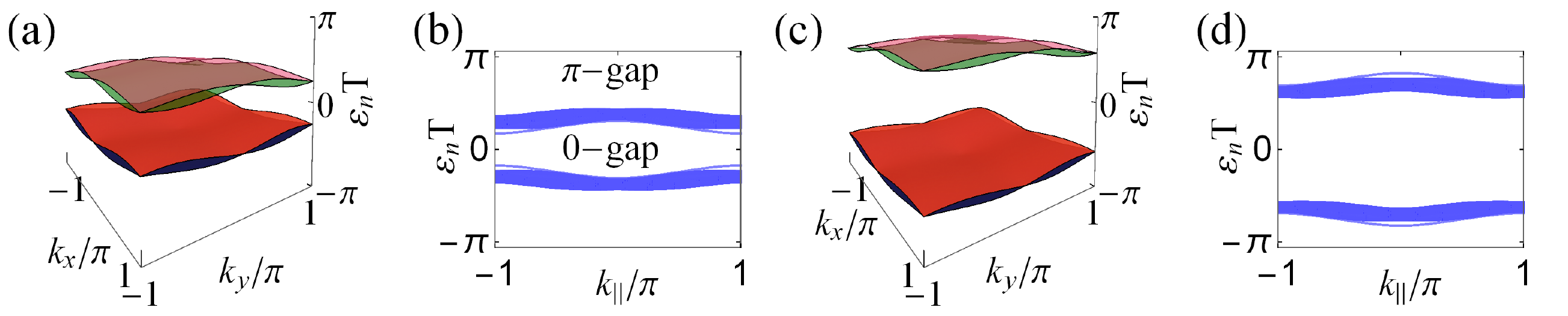}
\caption{Floquet spectra under PBC ((a)(c)) and stripe geometry ((b)(d)). For the stripe geometry, $k_{\parallel}$ denotes the lattice momentum parallel to the stripe. For (a)(b), $\phi_0/\pi=0.05$; For (c)(d), $\phi_0/\pi=0.45$. Other parameters are $t_x=t_y=\pi/T$.}
\label{figs1}
\end{figure}

Fig. \ref{figs1} (a)(c) depict their Floquet spectra under PBC. The  existence of both 0- and $\pi$- gap splits the bulk bands into two sets, with each set containing two degenerate bands at high-symmetry points in the 2D Brillouin zone (BZ). The two sets are each other's particle-hole partner. Using either set of the bulk Floquet bands, we can construct the Wannier bands using Wilson loop, similar to the static case. We will discuss this point in detail in the next part. Fig. \ref{figs1}(b)(d) depict the Floquet spectra under stripe geometry, which are fully gapped, without any in-gap modes along the 1D edges. This is a key difference from the first-order topological phase. Together with the Floquet spectra under OBC, the appearance of CMs is a salient feature of the higher-order topology of FQIs.

\begin{figure}[h!]
\centering
\includegraphics[width=5.6in]{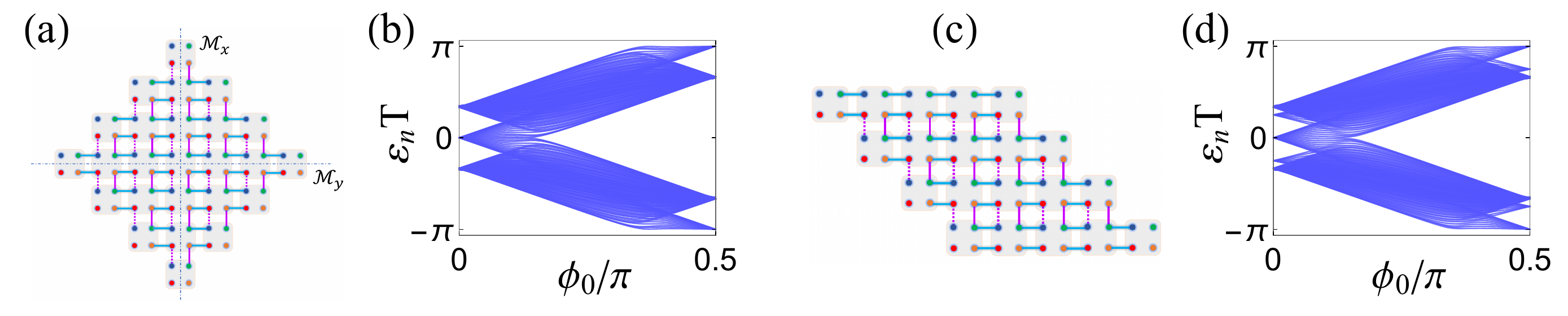}
\caption{(a)(c) Finite lattices with zigzag edges and (b)(d) their corresponding quasienergy spectra. In (a), the mirror symmetries $\mathcal{M}_x$ and $\mathcal{M}_y$ exchange different edges. The parameters are the same as Fig. 2 in the main text with $\phi_0=\frac{t_0T}{2\sqrt{2}}$.}\label{zig}
\end{figure}

The CMs are protected by mirror symmetries $\mathcal{M}_x$ and $\mathcal{M}_y$. In crystalline topological phases, boundaries of different orientations may host modes of different dispersions or robustness. The boundary/corner modes in the FQIs depend on the edge termination, and not all of them are robust. Two examples of zigzag edges and their corresponding Floquet spectra are illustrated in Fig. \ref{zig}(a)(c) and Fig. \ref{zig}(b)(d), respectively. In the limit of vanishing intra-cell coupling $t_0=0$, there are a great many degenerate edge/corner states at zero energy. However, these states are not protected. For finite intra-cell couplings $t_0$, the sites on the edge become coupled. Accordingly, the edge states acquire dispersion and move away from zero energy. One can numerically verify that there is no localized corner states at finite $t_0$. This is not surprising, because the four edges in Fig. \ref{zig}(a)(c) no longer respect both $\mathcal{M}_x$ and $\mathcal{M}_y$ symmetries, in contrast to the edges we considered in the main text. Consequently, the edge polarization is no longer quantized, and the corner states, if any, are no longer protected by mirror symmetries. 

\subsection{Nested Wilson loop approach}
The NWL approach provides an intuitive route towards the characterization of the hierarchical structure of bulk multipole moments and higher-order topology. Following previous literature \cite{hoti01,hoti02}, here we briefly review this method and show how to calculate the edge polarizations and bulk multipole moments for the static Bloch bands as well as Floquet bands. The multipole moments are quantized in the presence of mirror symmetries, yielding a $\mathbb{Z}_2$ classification of the topological multipole insulators. The following discussions apply to both static and periodically driven systems. For the latter, we need to replace the Bloch bands with Floquet bands.

The starting point is the Wilson loop, a unitary operator describing parallel transport of eigenstates along a closed path $\Gamma$ in the BZ. In thermodynamical limit, the Wilson loop is a path-ordered exponential:
\begin{eqnarray}
\mathcal{W}_{\Gamma}=P_{\Gamma}\exp(-i\oint_{\Gamma} d\bm k\cdot \bm A_{\bm k}),\label{wl}
\end{eqnarray}
where $\bm A_{\bm k}$ is the Berry connection of the Bloch/Floquet eigenstates $|\psi_{n\bm k}\rangle$, i.e., $\bm A_{\bm k}^{mn}=-i\langle \psi_{m\bm k}|\bm{\nabla}_{\bm k}| \psi_{n\bm k}\rangle$.

\textbf{(\textrm{I}) \textit{Topological quadrupole insulator in 2D}.}

\begin{figure}[h]
\centering
\includegraphics[width=4in]{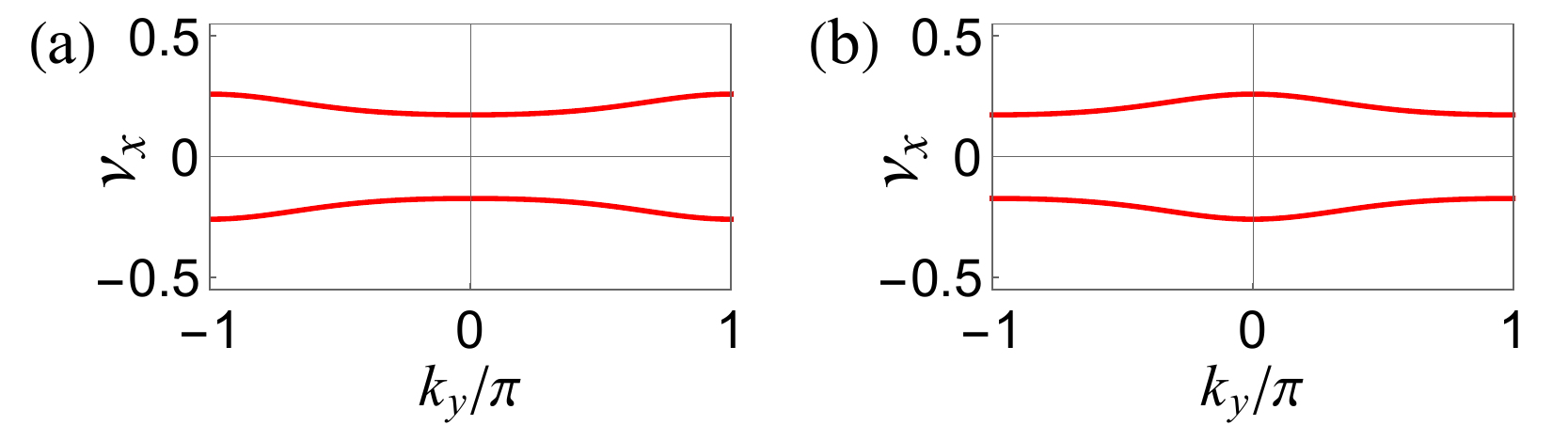}
\caption{Wannier bands $\nu_x$ with PBC for the two types of FQIs in driving protocol Eq. (4). (a) $\phi_0/\pi=0.1$; (b) $\phi_0/\pi=0.45$. Other parameters are $t_x=t_y=\pi/T$.}\label{figs2}
\end{figure}

We first consider the topological quadrupole insulator in 2D, with a minimal of four bulk bands \cite{hoti01,hoti02}. One can define $\mathcal{W}_{x\bm k}$ as the Wilson loop along a path parallel to $k_x$ in the 2D BZ with $\Gamma: (k_x,k_y)\rightarrow(k_x+2\pi,k_y)$. Here $\bm k$ is the base point of the Wilson loop (similarly for a path parallel to $k_y$, one can define $\mathcal{W}_{y\bm k}$). Compared to the static system with the band index $m,n$ in $\bm A_{\bm k}$ running over all the occupied bands, in our Floquet system, $m,n$ can be chosen in either set (with $-\pi/T<\varepsilon_{m\bm k},\varepsilon_{n\bm k}<0$ or with $0<\varepsilon_{m\bm k},\varepsilon_{n\bm k}<\pi/T$) of the bulk Floquet bands. With the identification that $\mathcal{W}_{x\bm k}$ is adiabatically connected to the physical Hamiltonian \cite{wannier} of the 1D edge: $\mathcal{W}_{x\bm k}=e^{-iH_{edge}(k_y)}$, it is well suited to characterize the boundary topology. Diagonalization of $\mathcal{W}_{x\bm k}$ yields
\begin{eqnarray}
\mathcal{W}_{x\bm k}|\nu_{x\bm k}^j\rangle=e^{i2\pi\nu_x^j(k_y)}|\nu_{x\bm k}^j\rangle.
\end{eqnarray}
Here $j=1,2$ and the $k_y$-dependence of the eigenphases has been explicitly written. $\nu^j_x(k_y)$ is proportional to the Wannier center (the position of electrons within the unit cell), forming the so-called Wannier band. Fig. \ref{figs2} depicts the two Wannier bands under PBC for both types of FQIs in our driving protocol Eq. (4). Here the base Floquet bands in Eq. (\ref{wl}) has been chosen to be inside $(-\pi,0)$. The Wannier bands are fully gapped over the entire range of $k_y\in[-\pi,\pi]$.

\begin{figure}[!t]
\centering
\includegraphics[width=5in]{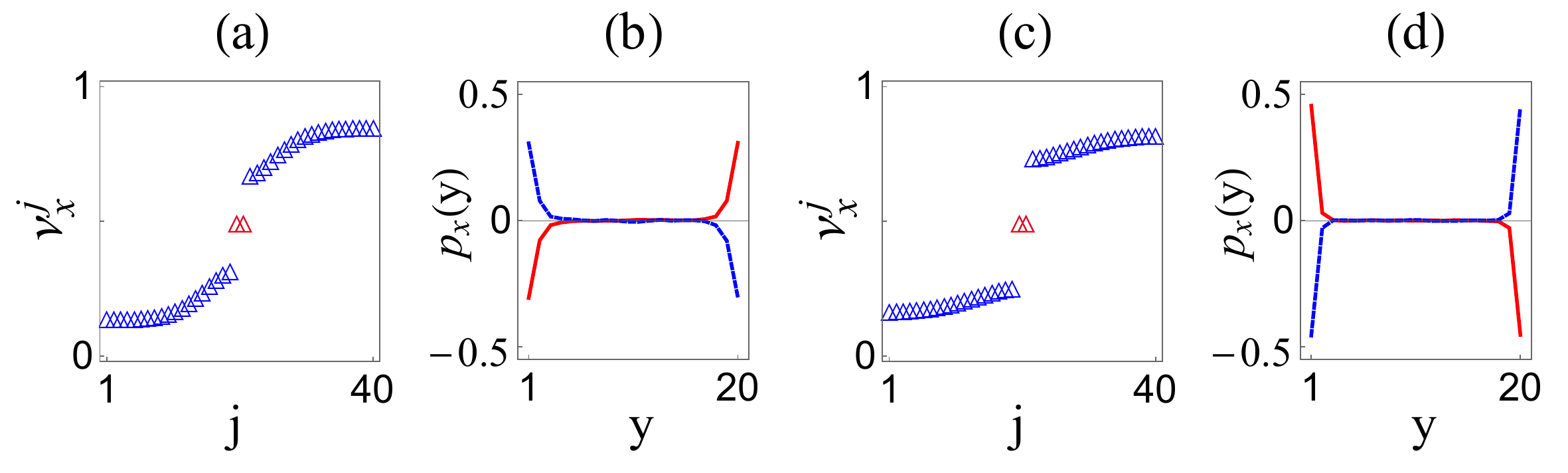}
\caption{Wannier-band characterizations of the two types of FQIs in model (4) of the main text. (a)(c) Wannier bands and (b)(d) tangential polarizations for a stripe of $L_y=20$, with $\phi_0/\pi=0.1$ for (a)(b) and $\phi_0/\pi=0.45$ for (b)(d).}
\label{figs3}
\end{figure}

As a manifestation of the underlying higher-order topology, the Wannier bands can carry their own topology, with the appearance of protected Wannier modes at the boundaries of the system. This can be confirmed by checking the Wannier bands with a stripe geometry. Numerically, by absorbing the lattice sites along the open direction ($\hat{y}$) into inner degrees of freedom, the Wannier bands can be calculated as
\begin{eqnarray}
\mathcal{W}_{x,k_x}|\nu_{x,k_x}^j\rangle=e^{i2\pi\nu_x^j}|\nu_{x,k_x}^j\rangle.
\end{eqnarray}
Here $j=1,2,...,2L_y$, $L_y$ is the number of unit cells along $y$. $\mathcal{W}_{x,k_x}$ denotes the Wilson loop with $k_x$ as its starting point. The resulting $\nu_x^j$ are depicted in Fig. \ref{figs3}(a) and \ref{figs3}(c) for the two types of FQIs identified in model (4) of the main text. In addition to the bulk values of $\nu_x^j$ (blue), there appear two in-gap modes (red) pinned at $1/2$ and localized at opposite boundaries. Utilizing the above Wannier bands, one can further calculate the tangential polarization, as an indication of the boundary topology. With the stripe geometry, the spatially resolved tangential polarization is defined as
\begin{eqnarray}
p_x(y)=\frac{1}{L_x}\sum_{j=1}^{2L_y}\sum_{k_x,\alpha}|\large[|\psi_{n,k_x}\rangle\large]^{y,\alpha}[|\nu_{x,k_x}^j\rangle]^n|^2\nu_x^j.
\end{eqnarray}
Here $|\psi_{n,k_x}\rangle$ is the Floquet eigenstate in the slab geometry. $[...]^n$ denotes the $n$-th component of the eigenvector, $y$ labels the unit cell and $\alpha$ labels the sites within the cell, and $n$ is summed over quasienergy bands with $\epsilon_n<0$. The spatial profile of $p_x(y)$ is shown in Fig. \ref{figs3}(b) and \ref{figs3}(d) for the two FQI phases, respectively. Although it vanishes in the bulk, the polarization $p_x(y)$ develops peaks at the two boundaries $y=1,20$. Moreover, we find the overall polarization is quantized to $1/2$ when $p_x(y)$ is integrated over half the slab $1\leq y<L_y/2$ (or $L_y/2<y\leq L_y$). The edge polarization vanishes after the topological phase transition, accompanied by the disappearance of CMs with full open boundaries. The protection by the bulk gap is a clear evidence that the edge polarization is caused by the bulk quadrupoles and not a pure edge property.

Based on the Wannier bands $|\nu_{x\bm k}^j\rangle~(j=1,2)$, one can construct the so-called NWL to characterize the bulk higher-order topology. Let us proceed by defining the Wannier-band subspace as
\begin{eqnarray}
|\mathcal{\omega}_{x,\bm k}^{j}\rangle =\sum_{n=1,2}|\psi_{n\bm k}\rangle[|\nu_{x\bm k}^{j}\rangle]^n.
\end{eqnarray}
$|\mathcal{\omega}_{x,\bm k}^{j}\rangle$ provides a natural splitting of the original pair of bands, which are degenerate at the high-symmetry points in the BZ for our driving system. Now in this single Wannier-band subspace, the NWL is the Wilson loop along $k_y$, i.e.,
\begin{eqnarray}
\tilde{\mathcal{W}}_{y,k_x}^{j}=F_{y,\bm k+L_y\delta \bm{k_y}}^{j}...F_{y,\bm k+\delta \bm{k_y}}^{j}F_{y,\bm k}^{j},
\end{eqnarray}
where $F_{y,\bm k}^{j}=\langle\mathcal{\omega}_{x,\bm k+\delta \bm {k_y}}^{j}|\mathcal{\omega}_{x,\bm k}^{j}\rangle$, with $\delta \bm{k_y}=(0,\frac{2\pi}{L_y})$. The associated polarization on the Wannier-band subspace is
\begin{eqnarray}
p_y^{\nu_x^{j}}=-\frac{i}{2\pi}\sum_{k_x}\log[\tilde{\mathcal{W}}_{y,k_x}^{j}],
\end{eqnarray}
which can be further represented as
\begin{eqnarray}
p_y^{\nu_x^{j}}=-\frac{1}{(2\pi)^2}\int_{BZ}\tilde{\mathcal{A}}_{y,\bm k}^{j}d^2 k,
\end{eqnarray}
in thermodynamical limit. Here $\tilde{\mathcal{A}}_{y,\bm k}^{j}=-i\langle\mathcal{\omega}_{x,\bm k}^{j}|\partial_{k_y}|\mathcal{\omega}_{x,\bm k}^{j}\rangle$ is the Berry potential over the Wannier-band subspace.

In the presence of mirror symmetries $\mathcal{M}_x$ and $\mathcal{M}_y$, the nested polarization $p_y^{\nu_x^{j}}$ and $p_x^{\nu_y^{j}}$ are quantized to be $0$ or $\frac{1}{2}$ \cite{hoti01,hoti02}, yielding a $\mathbb{Z}_2$ classification of the Wannier-band topology. Physically, when the effective edge Hamiltonian $H_{edge}(k_y)$ (for the edge parallel to $y$) is in a topological insulator phase, $p_y^{\nu_x^{j}}=\frac{1}{2}$; while for the trivial insulator phase, $p_y^{\nu_x^{j}}=0$. The above discussions and conclusions apply to the other edge. The topological quadrupole phase corresponds to $(p_y^{\nu_x^{j}},p_x^{\nu_y^{j}})=(\frac{1}{2},\frac{1}{2})$, with the $\mathbb{Z}_2$ quadrupole invariant in the main text being identified as
\begin{eqnarray}
\nu_0^F=4p_y^{\nu_x^{j}}p_x^{\nu_y^{j}}.\label{tqi}
\end{eqnarray}
$F$ in the notation indicates Floquet bands. Note that the tangential polarizations and quadrupole invariant are the same regardless of the choice of Wannier bands $j=1,2$. We can omit the index $j$ as in the main text. When $\nu_0^F=1$, the driving system is in FQI phase, featuring edge polarizations; while when $\nu_0^F=0$, the system is in a trivial insulator phase.

\textbf{(\textrm{II}) \textit{Topological octupole insulator in 3D}.}

The above procedure can be easily extended to the calculations of octupole moment in 3D. The minimal model of a topological octupole insulator requires eight bands \cite{hoti01,hoti02}. This can be understood from a microscopic point of view, where a bulk octupole can be regarded as two spatially separated quadrupoles of opposite signs. Naturally, the octupole moment of 3D materials manifest itself through the existence of surface-bound quadrupole moments. Let us take the surface perpendicular to $\hat{z}$ as example. Following previous discussions, the Wilson loop along a path parallel to $k_z$ (i.e., $\Gamma: (k_x,k_y,k_z)\rightarrow(k_x,k_y,k_z+2\pi)$) in the 3D BZ is denoted as $\mathcal{W}_{z\bm k}$. It is adiabatically connected to the surface Hamiltonian of the 3D material as
\begin{eqnarray}
\mathcal{W}_{z\bm k}=e^{-iH_{surf}(k_x,k_y)}.
\end{eqnarray}
Diagonalization of $\mathcal{W}_{z\bm k}$ gives
\begin{eqnarray}
\mathcal{W}_{z\bm k}|\nu_{z,\bm k}^j\rangle=e^{i2\pi\nu_{z}^j(k_x,k_y)}|\nu_{z,\bm k}^j\rangle,\label{wannier3d}
\end{eqnarray}
with $j=1,2,3,4$. The four eigenphase bands $\nu_{z}^j(k_x,k_y)$ can be further categorized into two gapped sectors, with each sector consisting of two overlapping bands. Note that the above phase bands resemble the bulk bands of a quadrupole insulator, with surface topology hidden inside. One can label each sector by $\pm$ respectively and rewrite Eq. (\ref{wannier3d}) as $\mathcal{W}_{z\bm k}|\nu_{z,\bm k}^{\pm,j}\rangle=e^{i2\pi\nu_{z}^{\pm,j}(k_x,k_y)}|\nu_{z,\bm k}^{\pm,j}\rangle$ for $j=1,2$. By choosing either sector (taking the $+$ sector as example), one can further construct the Wannier states
\begin{eqnarray}
|\omega_{z,\bm k}^{+,j}\rangle=\sum_{n=1}^4 |\psi_{n\bm k}\rangle[|\nu_{z,\bm k}^{+,j}\rangle]^n,
\end{eqnarray}
with $j=1,2$. Based on these Wannier states, the non-Abelian NWL along $k_y$ can be explicitly written as
\begin{eqnarray}
[{\tilde{\mathcal{W}}}^+_{y,\bm k}]^{j,j'}=\langle\omega^{+,j}_{z,\bm k+L_y\delta \bm k_y}|\omega^{+,r}_{z,\bm k+(L_y-1)\delta \bm k_y}\rangle\langle\omega^{+,r}_{z,\bm k+(L_y-1)\delta \bm k_y}|...|\omega^{+,s}_{z,\bm k+\delta \bm k_y}\rangle\langle\omega^{+,s}_{z,\bm k+\delta \bm k_y}|\omega^{+,j'}_{z,\bm k}\rangle,
\end{eqnarray}
with $\delta \bm k_y=(0,\frac{2\pi}{L_y},0)$, $j,r,...s,j'=1,2$. ${\tilde{\mathcal{W}}}^+_{y,\bm k}$ is a $2\times 2$ matrix. The summation over repeated index is assumed. Physically, ${\tilde{\mathcal{W}}}^+_{y,\bm k}$ is adiabatically connected to the 1D hinge Hamiltonian $H_{hinge}(k_x)$ (the hinger shared by $\hat{x}\hat{y}$ and $\hat{x}\hat{z}$ plane) as
\begin{eqnarray}
{\tilde{\mathcal{W}}}^+_{y,\bm k}=e^{-iH_{hinge}(k_x)}.
\end{eqnarray}
All the above procedures finally bring us to the boundary of the boundary. In a topological octupole phase, $H_{hinge}(k_x)$ should have the same topology as 1D topological insulator. Further diagonalizing ${\tilde{\mathcal{W}}}^+_{y,\bm k}$, i.e.,
\begin{eqnarray}
{\tilde{\mathcal{W}}}^+_{y,\bm k}|\eta_{y,\bm k}^m\rangle=e^{i2\pi\eta_y^m}|\eta_{y,\bm k}^m\rangle,
\end{eqnarray}
yields two gapped Wannier bands $\eta_y^m~(m=1,2)$. We define the Wannier-sector $|\omega_{y,\bm k}^{m}\rangle$ by choosing either one as
\begin{eqnarray}
|\omega_{y,\bm k}^{m}\rangle=\sum_{n=1}^2 |\psi_{n\bm k}\rangle[|\eta_{y,\bm k}^{m}\rangle]^n.
\end{eqnarray}
The NWL and associated Wannier-sector polarization along $\hat{x}$ can then be calculated as
\begin{eqnarray}
{\tilde{\mathcal{W}}}^m_{x,\bm k}&=&\langle\omega^{m}_{y,\bm k+L_x\delta \bm k_x}|\omega^{m}_{y,\bm k+(L_x-1)\delta \bm k_x}\rangle\langle\omega^{m}_{y,\bm k+(L_x-1)\delta \bm k_x}|...|\omega^{m}_{y,\bm k+\delta \bm k_x}\rangle\langle\omega^{m}_{y,\bm k+\delta \bm k_x}|\omega^{m}_{y,\bm k}\rangle,\\
p_x^m&=&-\frac{i}{2\pi}\frac{1}{L_yL_z}\sum_{k_y,k_z}\log[{\tilde{\mathcal{W}}}^m_{x,\bm k}].
\end{eqnarray}
Physically, if $H_{hinge}(k_x)$ is in a 1D topological/trivial insulator phase, $p_x^m=\frac{1}{2}/0$. By tracing back the whole procedure, the non-trivial nested polarization of $p_x^m$ is a consequence of the 2D quadrupole topology of $H_{surf}(k_x,k_y)$ as well as the 3D octupole topology of  $H(\bm k)$. In the above calculations, the order of Wilson loop nesting $\mathcal{W}_z\rightarrow\mathcal{W}_y\rightarrow\mathcal{W}_x$ and the choice of Wannier bands are arbitrary. Similar discussions can be performed for other hinges and Wannier-sector polarization $p_y^m$ and $p_z^m$.

In the presence of mirror symmetries $\mathcal{M}_x$, $\mathcal{M}_y$, $\mathcal{M}_z$, the octupole moment and Wannier-sector polarizations are quantized to be $\frac{1}{2}$ or $0$ \cite{hoti01,hoti02}, yielding a $\mathbb{Z}_2$ classification of the octupole phase. We note that each surface of a topological octupole insulator is a topological quadrupole insulator. With the characterization of the latter on hand [see Eq. (\ref{tqi})], it is natural to define the $\mathbb{Z}_2$ invariant as
\begin{eqnarray}
\nu_0^F=8p_x^m p_y^m p_z^m,
\end{eqnarray}
with $F$ refering to the Floquet bands. When $\nu_0^F=1$, the system is in Floquet topological octupole phase, featuring surface quadrupoles as well as hinge polarizations; while when $\nu_0^F=0$, the system is in a trivial insulator phase.

\subsection{$\mathbb{Z}_2$ invariants with mirror symmetries}
In the main text, we have introduced two $\mathbb{Z}_2$ invariants $\nu_0$ and $\nu_{\pi}$ as follows:
\begin{eqnarray}
\nu_\pi=n_{\pi} \bmod 2;~~~\nu_0=(n_0+\nu_0^F) \bmod 2,\label{topoinvas}
\end{eqnarray}
to characterize the CMs in each quasienergy gap. Here $\nu_0^F$ is the $\mathbb{Z}_2$ invariant of the Floquet Hamiltonian $H_F$ defined by NWL and $n_{0,\pi}$ counts the dynamical Weyl charges (gap-closings) in the phase bands of return map $\tilde{U}(t)$. We note that the topological invariants $\nu_{0,\pi}$ are general, and work for all FHOTIs protected by mirror symmetries, not just for the model or specific examples given in the main text. 

\begin{figure}[!h]
\centering
\includegraphics[width=4.5in]{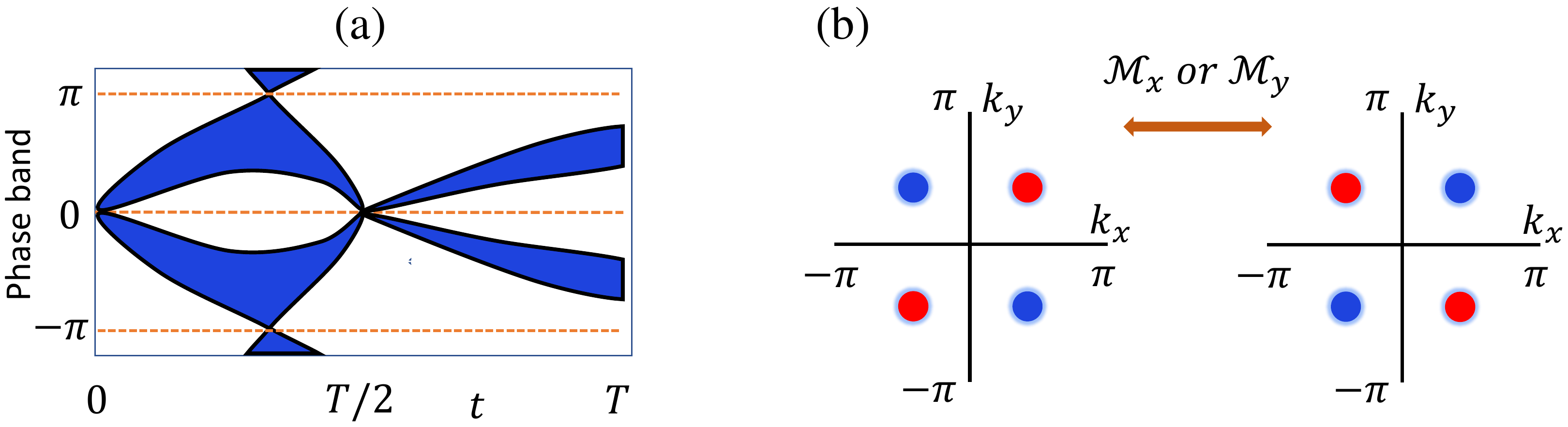}
\caption{(a) Decomposition of unitary evolution $U(t)$ into return map (or unitary loop) $\tilde{U}(t)$ and constant evolution with $H_F$. (b) Equivalence of the two quadrupole patterns (+-+-) and (-+-+) in the BZ under mirror symmetries. The red/blue dots represent $+/-$ Weyl charges.}
\label{figz2}
\end{figure}

The $\mathbb{Z}_2$ invariants $\nu_{0,\pi}$ defined above follow the standard constructions of Floquet topological invariants \cite{fclass1,fclass2,fclass3}. For Floquet system, the full evolution is decomposed into two parts. Mathematically, this decomposition is homotopic to a two-step evolution \cite{fclass2}: $\tilde{U}(t)$ (also called unitary loop) followed by a constant evolution $H_F$, as illustrated in Fig. \ref{figz2}(a). The loop unitary $\tilde{U}(t)$ may generate CMs. As the eigenvalues of $H_F$ (for fully gapped system) have magnitude strictly less than $\frac{\pi}{T}$, the $\pi$-CMs which are determined by gap closings of $U(t)$ in the whole time evolution, are attributed to the $\pi$-gap closings of $\tilde{U}(t)$; while the 0-CMs are determined by both $H_F$ and the loop unitary: if no $0$-gap closing happens for $\tilde{U}(t)$, the 0-CMs are solely determined by the topology of $H_F$, however, if $0$-gap closings happen, the Floquet topological invariant $\nu_0$ should include the additional contributions from $\tilde{U}(t)$. 

Next, we show the necessity of taking module 2 in Eq. (\ref{topoinvas}) due to mirror symmetries $\mathcal{M}_x$ and $\mathcal{M}_y$. This renders a $\mathbb{Z}_2\times\mathbb{Z}_2$ classification for FHOTIs due to both $0$- and $\pi$-gaps. Under mirror reflections, we have
\begin{eqnarray}
\mathcal{M}_x \tilde{U}(k_x,k_y,t)\mathcal{M}_x^{-1}&=&\tilde{U}(-k_x,k_y,t),\notag\\
\mathcal{M}_y \tilde{U}(k_x,k_y,t)\mathcal{M}_y^{-1}&=&\tilde{U}(k_x,-k_y,t).
\end{eqnarray}
Hence the Weyl charges always come in quartets and form a quadrupole pattern in the BZ at a specific time instant. To see that, we consider for example a Weyl charge at $(k_{x}^0, k_{y}^0,t^0)$ and $\tilde{U}$ takes the form $a_x\delta k_x\sigma_x+a_y\delta k_y\sigma_y+a_t\delta t\sigma_z$ in its neighborhood (the other two bands are energetically far away and irrelevant). Then near $(-k_x^0,k_y^0,t^0)$, $\tilde{U}$ should take the form $-a_x\delta k_x\sigma_x+a_y\delta k_y\sigma_y+a_t\delta t\sigma_z$. These two Weyl charges have opposite signs. Similarly arguments apply to Weyl charges at $(k_x^0,-k_y^0,t^0)$ and $(-k_x^0,-k_y^0,t_0)$. The charge distributions from the first to fourth quadrant of BZ can be either $(+-+-)$ or $(-+-+)$ as illustrated in Fig. \ref{figz2}(b). Now we show these two patterns are in fact equivalent to each other. Mathematically, we can find a continuous path (i.e., without introducing any new gap-closings or -openings) to connect the two patterns. Using mirror symmetry $\mathcal{M}_x$ (or $\mathcal{M}_y$), the continuous path (parameterized by $\theta$) can be constructed as: 
\begin{eqnarray}
\tilde{U}(k_x,k_y,t,\theta)=e^{\theta\mathcal{M}_x}\tilde{U}(k_x,k_y,t)e^{-\theta\mathcal{M}_x}.
\end{eqnarray}
Obviously $\tilde{U}(k_x,k_y,t,\theta=0)=\tilde{U}(k_x,k_y,t)$; $\tilde{U}(k_x,k_y,t,\theta=\frac{\pi}{2})=\mathcal{M}_x\tilde{U}(k_x,k_y,t)\mathcal{M}_x^{-1}=\tilde{U}(-k_x,k_y,t)$, where $\mathcal{M}_x^2=-I$ is used. Note $\tilde{U}(k_x,k_y,t)$ and $\tilde{U}(-k_x,k_y,t)$ have opposite quadrupole patterns and along the whole path $0\leq\theta\leq\frac{\pi}{2}$, the phase bands keep unchanged.

Therefore we can conclude the two patterns are equivalent and the Weyl charge in each quadrant of BZ is only defined module 2. It is worth to mention the difference from traditional Weyl points without any symmetry constraints. In Weyl semimetals, only two Weyl points with opposite charges can annihilate each other and open a gap when brought together. The equivalence of the two quadrupole patterns indicates that any two of them can annihilate each other and open a gap when brought together, even when they have the same charge in the first quadrant of BZ. (If this is the case, we can continuously deform one of them to its opposite sign and annihilate the other in each quadrant as in the Weyl semimetal case.) 

\subsection{Phase-band touchings for the anomalous case}
\begin{figure}[h]
\centering
\includegraphics[width=5.5in]{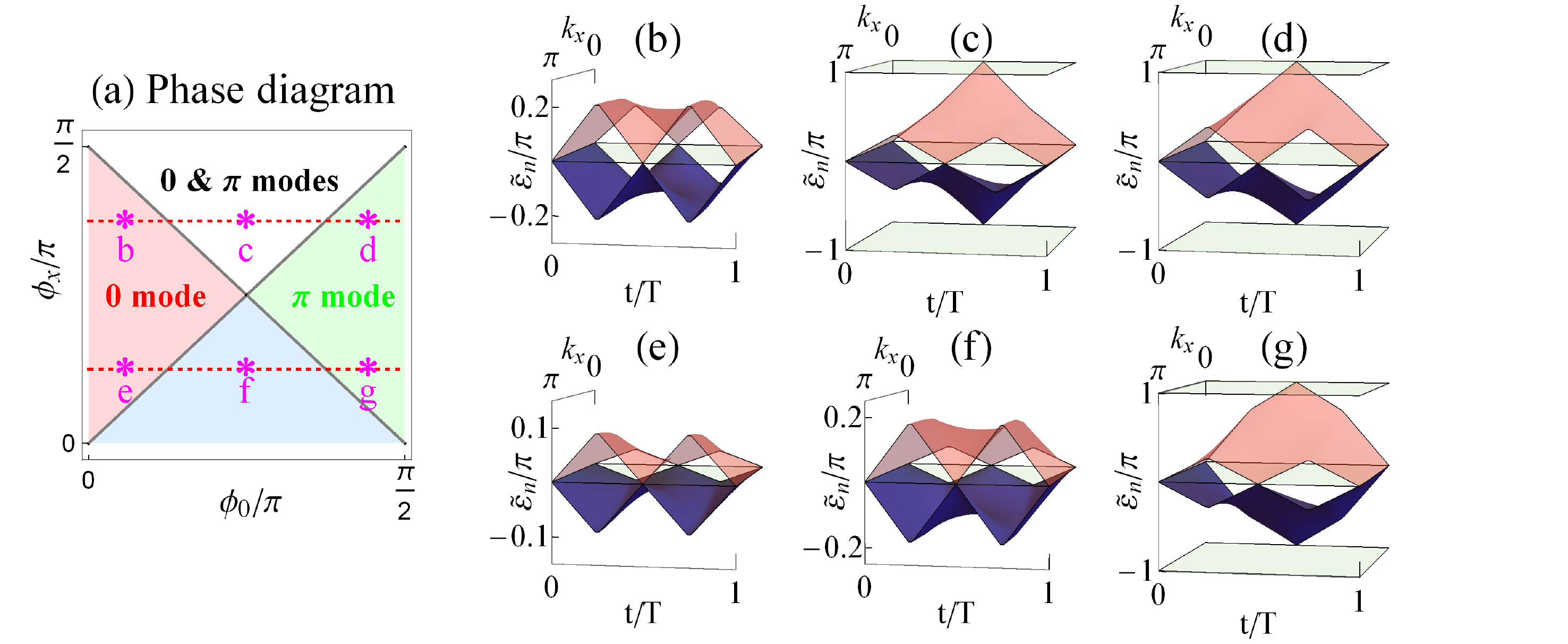}
\caption{Phase-band characterizations of the Floquet topological phases of driving protocol Eq. (\ref{prot2}). (a) Phase diagram in the $\phi_0-\phi_x$ plane. The four phases with different types of CMs are labeled by different colors. (b)-(g) Phase bands of $\tilde{U}(t)$ projected onto the $k_x = k_y$ line for the chosen parameters in (a) (labeled by $*$). Each band is two-fold degenerate.}
\label{pbano}
\end{figure}

In this part, we demonstrate the phase bands of $\tilde{U}(t)$ for Floquet driving ($h_0, h_{x,y}$ are the same as Eq. (4)):
\begin{eqnarray}
0<t<\frac{T}{4},~H(t)=h_0;~~~\frac{T}{4}<t<\frac{3T}{4},~H(t)=h_x+h_y;~~~\frac{3T}{4}<t<T, ~H(t)=h_0.\label{prot2}
\end{eqnarray}
The phase boundaries for the special case $t_x=t_y$ are determined by \cite{fhoti1} ($\phi_0\equiv\frac{t_0T}{2\sqrt{2}}$, $\phi_x\equiv\frac{t_xT}{2\sqrt{2}}$)
\begin{eqnarray}
\phi_0=\pm \phi_x+\frac{\mathcal{N}\pi}{2},~\mathcal{N}\in\mathbb{Z}.
\end{eqnarray}
The phase diagram is illustrated in Fig. \ref{pbano}(a). Totally there are four distinct Floquet phases, featured by the appearance of different CMs. Let us examine their phase bands in detail.

For the two FQI phases with either 0- or $\pi$-CMs, $\nu_0^F=1$, indicating the existence of bulk quadrupole moment. For the 0-CM case [Fig. \ref{pbano}(b)(e)], the phase bands touch $\tilde{\varepsilon}=0$ twice at $\bm k=(0,0)$ and $\bm k=(\pi,\pi)$ (there is no touching with quasienergy zone edge), yielding $(\nu_0,\nu_{\pi})=(1,0)$ from Eq. (9); while for the $\pi$-CM case [Fig. \ref{pbano}(d)(g)], the phase bands touch both zero and the zone edge once, yielding $(\nu_0,\nu_{\pi})=(0,1)$. For the anomalous phase with both 0- and $\pi$-CMs and trivial phase without any CMs, $\nu_0^F=0$. The phase bands of the former [Fig. \ref{pbano}(c)] touch both zero and the zone edge once, yielding $(\nu_0,\nu_{\pi})=(1,1)$. The phase bands for the latter [Fig. \ref{pbano}(f)] touch zero twice, yielding $(\nu_0,\nu_{\pi})=(0,0)$.

\subsection{Phase boundaries and phase-band characterizations of Floquet octupole insulators}
In this part, we derive the phase boundaries of different Floquet phases [see Eq. (10) in the main text] and demonstrate their phase-band characterizations in 4D $(\bm k,t)$-space. For convenience, we focus on the $t_x=t_y=t_z$ case.  The stroboscopic evolution operator $U(T)=f(\bm k)+ig(\bm k)$ can be represented as ($\phi_0\equiv\frac{\sqrt{3}t_0T}{4}$, $\phi_x\equiv\frac{\sqrt{3}t_xT}{4}$)
\begin{eqnarray}
U(T)&=&e^{-i \frac{h_0T}{4}}e^{-i \frac{(h_x+h_y+h_z)T}{2}}e^{-i \frac{h_0T}{4}}\nonumber\\
=&&[\cos\phi_0-i\sin\phi_0\frac{h_0}{\sqrt{3}t_0}][\cos2\phi_x-i\sin2\phi_x\frac{h_x+h_y+h_z}{\sqrt{3}t_x}][\cos\phi_0-i\sin\phi_0\frac{h_0}{\sqrt{3}t_0}].
\end{eqnarray}
The real part of $U(T)$ determines the phase boundaries, which can be explicitly calculated as
\begin{eqnarray}
f(\bm k)=\cos2\phi_0\cos2\phi_x-\frac{\sin2\phi_0\sin2\phi_x}{3}(\cos k_x+\cos k_y+\cos k_z).
\end{eqnarray}
The topological phase transitions are due to gap closings located at $0$ and $\frac{\pi}{T}$, i.e., when the eigenvalues of $f(\bm k)$ take $\pm 1$. The 0-gap closing condition (dictating the appearance/disappearance of 0-CMs) is then
\begin{eqnarray}
\cos2\phi_0\cos2\phi_x\pm\sin2\phi_0\sin2\phi_x=1.
\end{eqnarray}
with solutions
\begin{eqnarray}
\phi_0\pm\phi_x=\mathcal{N}\pi,~~\mathcal{N}\in \mathbb{Z}.
\end{eqnarray}
Similarly, The $\pi$-gap closing condition (dictating the appearance/disappearance of $\pi$-CMs) is
\begin{eqnarray}
\cos2\phi_0\cos2\phi_x\pm\sin2\phi_0\sin2\phi_x=-1.
\end{eqnarray}
with solutions
\begin{eqnarray}
\phi_0\pm\phi_x=(\mathcal{N}+\frac{1}{2})\pi,~~\mathcal{N}\in \mathbb{Z}.
\end{eqnarray}
The Floquet phase diagram has already been illustrated in Fig. 4(b) in the main text, in consistent with the above boundary conditions.
\begin{figure}[h]
\centering
\includegraphics[width=7in]{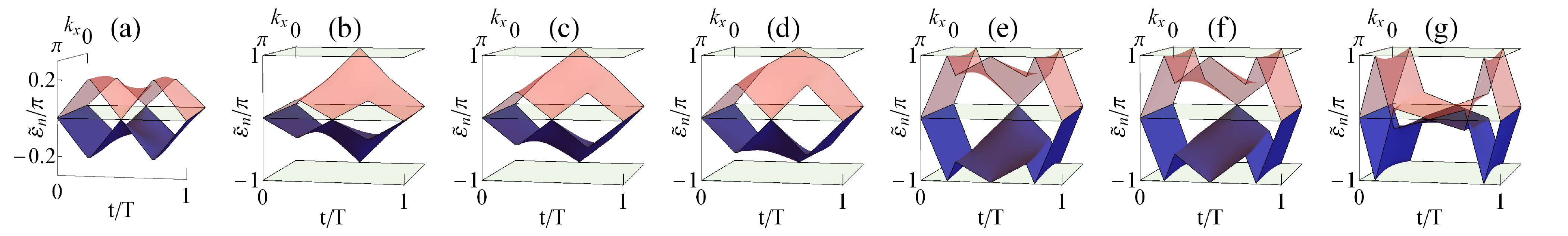}
\caption{Phase bands projected onto the $k_x=k_y=k_z$ line for different types of FOIs. From left to right, (a) $\phi_0/\pi=0.05$; (b) $\phi_0/\pi=0.25$; (c) $\phi_0/\pi=0.5$; (d) $\phi_0/\pi=0.75$; (e) $\phi_0/\pi=1.25$; (f) $\phi_0/\pi=1.5$; (g) $\phi_0/\pi=1.95$. Each band-touching point is eight-fold degenerate. $\phi_x/\pi=\frac{3}{8}$.}
\label{figsfoi}
\end{figure}

Now we show the dynamical characterizations of different FOIs from the phase bands. The full topology comes from both the Floquet Hamiltonian $H_F(\bm k)$ and the phase-band singularities. First, we calculate $\nu_0^F$ using the NWL approach introduced above and find: for the FOI with either 0- or $\pi$-CMs, $\nu_0^F=1$; for the FOI with both types of CMs and trivial case, $\nu_0^F=0$. We note that the third phase is specific to driven systems and dub it as anomalous FOI. For the seven phases in Fig. \ref{figsfoi}(a)-(g), $\nu_0^F=1,0,1,0,0,1,1$ from left to right. The rest is to determine the band touchings in the phase bands of $\tilde{U}(t)$. The relevant touchings happen at four diagonal lines $k_x=\pm k_y=\pm k_z$. We take $k_x=k_y=k_z$ line as an example. The phase bands projected onto this line for different phases are illustrated in Fig. \ref{figsfoi}. Note that due to additional symmetries, each phase band is four-fold degenerate. The relevant band-touching points are eight-fold degenerate. From left to right, the band touchings with zero quasienergy are $2,1,1,1,1,1,4$ times and with quasienergy zone edge are $0,1,1,1,5,5,4$ in the 4D $(\bm k,t)$-space, yielding $\nu_0=1,1,0,1,1,0,1$ and $\nu_{\pi}=0,1,1,1,1,1,0$ for the seven phases. These results fully agree with the appearance of CMs and phase diagram [see Fig. 4(b) in the main text].

\subsection{Evolution of corner modes and entanglement entropy}
\begin{figure}[h]
\centering
\includegraphics[width=3.3in]{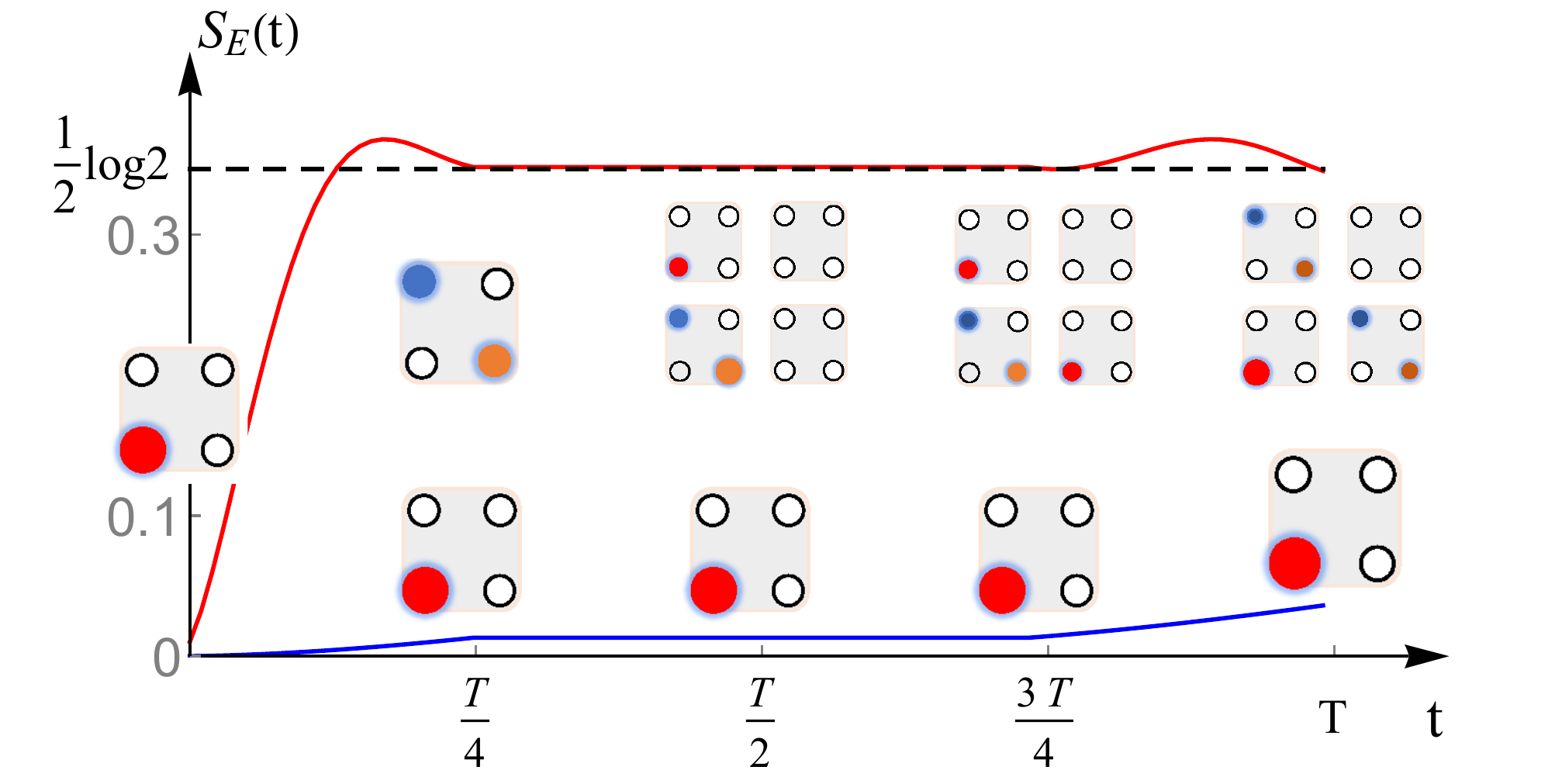}
\caption{Time-dependent entanglement entropy $S_E(t)$ for FQIs with 0-CMs (blue line) and $\pi$-CMs (red line), respectively. $t_x=t_y=\pi/T$. $\phi_0/\pi=0.05$ for 0-CM case and $\phi_0/\pi=0.45$ for $\pi$-CM case. The insets schematically plot the particle distributions at four intermediate time instant $t=\frac{T}{4}$, $\frac{T}{2}$, $\frac{3T}{4}$, and $T$.}
\label{ee}
\end{figure}

The evolution of CMs can reveal different dynamics of the two types of FQIs [see Fig. 2(a)] in the driving protocol Eq. (4). Here we provide another perspective from the entanglement entropy (EE) and consider the dynamical evolution of a single particle initially localized at the corner with wave function $\psi(t=0)\rangle=|1\rangle_{ld}$. The time-evolved wave function at time $t$ is given by
\begin{eqnarray}
|\psi(t)\rangle=U(t)|\psi(t=0)\rangle.
\end{eqnarray}
By cutting along the diagonal line of the square lattice (the lattice is now split into two parts A and B), the time-dependent EE for subsystem A is defined as
\begin{eqnarray}
S_E(t)=-\textrm{Tr}[\rho_{\textrm{A}}(t)\log\rho_{\textrm{A}}(t)],
\end{eqnarray}
where $\rho_A(t)$ is the reduced density matrix for subsystem A. Formally, the EE of a non-interacting fermion state $|\psi(t)\rangle$ can be calculated using the correlation functions as
\begin{eqnarray}
S_E(t)=-\sum_n C_n(t)\log C_n(t).
\end{eqnarray}
Here $C_n(t)$ are the eigenvalues of the correlation matrices defined as
\begin{eqnarray}
C_{is,js'}=\langle\psi(t)|c_{is}^{\dag}c_{js'}|\psi(t)\rangle.
\end{eqnarray}
Now let us examine the corner dynamics, as sketched in Fig. \ref{ee}. For the FQI with 0-CMs, the initial particle (red) will basically stay at the left-down corner in the whole period. It will eventually disperse into the bulk after a long time due to the non-vanishing overlapping of the 0-CM wave function with the bulk. While for FQI with $\pi$-CMs, the particle will first move to its nearest two sites $2$ (orange), $4$ (blue) in the first step, and disperse into the bulk in the second and third step. Such difference in dynamics is governed by the time-dependent EE. We can see $S_E(t)$ stays almost at zero for the FQI with 0-CMs; while for the FQI with $\pi$-CMs, it will saturate to $\frac{\log 2}{2}$.

\begin{thebibliography}{99}
\bibitem{footnote} The four ``Weyl'' charges may merge together by additional symmetries and form a multi-fold degenerate band singularities at the high-symmetry point. In this case, we can directly count $n_{0}$, $n_{\pi}$ as the times of multi-fold band touchings in Eq. (\ref{topoinva}). A multi-fold band touching will split into four ``Weyl'' charges by small perturbations preserving mirror symmetries.
\end{thebibliography}
\end{document}